\documentclass[useAMS,usenatbib,psfig]{mn2e}

\usepackage{amssymb}
\usepackage{times}
\usepackage{color}
\usepackage{ulem}
\usepackage{graphicx}

\title[Jet production in X-ray binaries and AGNs] {Jet production in black-hole X-ray binaries and active galactic nuclei:
mass feeding and advection of magnetic fields}
\author[X. Cao \& D. Lai]
{Xinwu Cao$^{1,2,3}$ and Dong Lai$^{4}$\\
$^1$ Shanghai Astronomical Observatory, Chinese Academy of Sciences,
80 Nandan Road, Shanghai, 200030, China; E-mail: cxw@shao.ac.cn\\
$^2$ Key Laboratory of Radio Astronomy, Chinese Academy of Sciences,
210008 Nanjing, China\\
$^3$ University of Chinese Academy of Sciences, Beijing 100049,
China\\
$^4$ Cornell Center for Astrophysics and
Planetary Science, Department of Astronomy, Cornell University,
Ithaca, NY 14853; E-mail: dong@astro.cornell.edu}

\date{Accepted 2019 February 23. Received 2019 February 21; in original form 2017 December 23}

\pagerange{\pageref{firstpage}--\pageref{lastpage}} \pubyear{}

\begin{document}

\maketitle \label{firstpage}

\begin{abstract}
Relativistic jets are observed only in the low/hard and intermediate
states of X-ray binaries (XRBs), and are switched off in the thermal
state, but they appear to be present in both low-luminosity and
luminous active galactic nuclei (AGNs). It is widely believed that strong large-scale magnetic fields
is a crucial ingredient in jet production; such fields can be attained only through efficient advection from
the outer disc. We suggest that geometrically thin accretion discs
with magnetic outflows are present in luminous radio-loud AGNs; this
is likely because the interstellar medium provides both mass and
sufficient magnetic flux to the outer disc. Most angular momentum of
such disc is removed by the outflows, and the radial velocity of the
disc is significantly increased compared to viscous drift velocity.
This facilitates efficient magnetic field advection through the disc
to produce a strong field near the black hole in luminous AGNs,
which helps launch relativistic jets. In XRBs, the magnetic fields
of the gas from companion stars are too weak to drive outflows from
outer discs. Jets are therefore switched off in the thermal state
due to inefficient magnetic field advection in the disc.
\end{abstract}


\section{Introduction}\label{intro}

Accretion discs are present both in black hole X-ray binaries (XRBs)
and active galactic nuclei (AGNs), with the black hole (BH) masses
ranging from $\la 10M_\odot$ in XRBs to $\ga
10^9M_\odot$ in AGNs.  In XRBs, discs are fed by mass transfer
from a companion star, while in AGNs circumnuclear gas in the host galaxy
is accreted onto a central suppermassive BH. The physics of accretion discs in these two types of sources is almost
same, and observations show that they indeed share many common properties.

Although both XRBs and AGNs produce relativistic jets, a direct maping
between the disc-jet phenomenologies of XRBs and AGNs is unclear
\citep*[e.g.,][]{2006MNRAS.372.1366K,2010LNP...794..115F}.  In XRBs,
X-ray outbursts are known to occur in cycles of several distinct
spectral states \citep*[][]{2005A&A...440..207B}. Steady compact jets
are always associated with the low/hard state, characterized by hard
power law in the X-ray spectrum.  Powerful episodic jets are produced
in the intermediate state, containing both hard power law component
and soft spectral component.  In the high-soft (thermal) state, radio
jets are switched off
\citep*[][]{1999ApJ...519L.165F,2001ApJ...554...43C,2003MNRAS.343L..99F,2003MNRAS.344...60G,2004ARA&A..42..317F,2004MNRAS.355.1105F},
although non-relativistic winds/outflows in the thermal state have
been observed in some X-ray binaries
\citep*[][]{2009Natur.458..481N,2009ApJ...695..888U,2012ApJ...746L..20K,2012ApJ...759L...6M}.
The origin of such winds/outflows is unclear
\citep*[][]{2009Natur.458..481N}.  It is believed that the spectral
states of BH XRBs correspond to different accretion modes. In the
low/hard state, the standard thin disc is truncated at a large
distance from the BH, and is replaced in the inner region by a
geometrically thick, radiatively inefficient flow
\citep*[e.g.,][]{1997ApJ...489..865E,2000A&A...354L..67M,2004MNRAS.351..791Z,2008ApJ...682..212W,2010LNP...794...53B}.

In the thermal state, the thin disc extends to the innermost stable orbit
around the BH.  It is clear that the production of relativistic jets
is regulated by accretion mode.
\footnote{We note that this ``standard" view remains under debate.
For example, there is evidence of broad Fe K$\alpha$ lines in the low-hard state of some X-ray
binaries \citep*[][]{2010MNRAS.402..836R,2016A&A...589A..14D}. This has been long-standing debate in the AGN/XRBs community (i.e., whether there are Fe alpha lines in the hard state of XRBs, or whether the detected lines originate from a thermal inner disc). If these lines indeed originate from an inner thermal disc in the hard state, it would contradict the standard picture held by many researchers in the field. Clearly, it is not our intention to resolve this debate in our paper.}
The role of magnetic field in state transition and jet formation in XRBs has been explored in some previous works \citep*[e.g.,][]{2012A&A...538A...5K,2015A&A...574A.133K,2016RAA....16...48L,2016ApJ...817...71C}. The phenomenology of state transition and jet production in AGNs is
less clear. Low-luminosity AGNs may contain radiatively inefficient
flows, which correspond to the low/hard state of XRBs, while
standard thin accretion discs are present in luminous AGNs (thermal
state)
\citep*[][]{1999ApJ...519...89C,1999MNRAS.303L..11Z,2005ApJ...625..667J,2005Ap&SS.300..219H,2007MNRAS.381.1235V}.
The observed jets in AGNs always exhibit a flat-spectrum radio core,
which are similar to the steady jets observed in the low/hard state
of XRBs. {Otherwise, one may expect that a compact radio core is
present only in a fraction of radio AGNs, if they contain episodic
jets as those in the intermediate state of XRBs.} By analogy with
XRBs, one might expect that relativistic jets preferentially appear
in low-luminosity AGNs. However, several lines of evidence suggest
that jets are also produced in luminous AGNs (analogy of thermal
state of XRBs). (i) The Eddington ratios of many radio-loud quasars
are $\sim 0.1-1$
\citep*[e.g.,][]{2007ApJ...658..815S,2009MNRAS.398.1905W}. For some
radio-loud narrow-line Seyfert 1 galaxies, the Eddington ratios can
be as high as $\ga 1$
\citep*[][]{2003ApJ...584..147Z,2006ApJ...639..710K,2006AJ....132..531K,2008ApJ...685..801Y,2013MNRAS.431..210C}.
Relativistic jets have been observed in radio-loud narrow-line
Seyfert 1 galaxies
\citep*[][]{2007PASJ...59..703D,2010AJ....139.2612G}. These indicate
that both radio-loud quasars and narrow-line Seyfert 1 galaxies with
jets contain radiatively efficient accretion discs, either standard
thin discs or slim discs
\citep*[][]{1973A&A....24..337S,1988ApJ...332..646A}. (ii) FR I
radio galaxies have lower radio power than FR II galaxies, and they
have different radio morphologies \citep*[][]{1974MNRAS.167P..31F}.
Using a sample of FR galaxies, \citet{2001A&A...379L...1G} suggested
that FR I galaxies can be separated from FR II galaxies by the
Eddington ratio $\sim 0.001-0.01$, which implies different accretion
modes in these two types of sources. (iii) In the unification scheme
for radio-loud AGNs, BL Lacertae (BL Lac) objects are believed to be
FR I radio galaxies with the jets aligned to our line of sight,
while radio quasars correspond to FR II galaxies viewed at small
angles with respect to the jet orientation. {The Eddington ratio
distributions of BL Lac objects/quasars are separated} \citep*[see Figure
2 in][]{2009ApJ...694L.107X}, which implies that radio quasars may
contain radiatively efficient accretion discs, while radiatively
inefficient flows may be in BL Lac objects. {For most BL Lac
objects in the Xu et al.~(2009) sample, the BH masses are estimated from
the host galaxy luminosities by using the $M_{\rm bh}-L_R$ relation
derived by \citet{2004MNRAS.352.1390M}. For a few BL Lac objects
with measured stellar dispersion velocity $\sigma$, the BH
masses are estimated with the empirical $M_{\rm bh}-\sigma$
relation. The bolometric luminosity of BL Lac objects is derived
from their broad/narrow line emission, because the continuum
emission may be dominated by Doppler beamed jet emission
\citep*[see][for the details]{2009ApJ...694L.107X}.} (iv) A big blue
bump in continuum spectra has been detected in some radio quasars,
which is taken as a characteristics of standard thin accretion discs
surrounding supermassive BHs \citep*[][]{2009MNRAS.400.1521J}. (v)
The observation of a broad ionized Fe K$\alpha$ line shows that a
thin accretion disc extends to the innermost region near the black
hole in the broad-line radio galaxy 4C~74.26
\citep*[][]{2005ApJ...622L..97B}. Overall, it appears that jet
formation in AGNs is not strongly regulated with the accretion mode,
contrary to the case of XRBs.

The production of relativitic jets in accreting BH systems (XRBs and
AGNs) relies on extracting the rotational energies of the inner
accretion discs and/or BHs through large-scale magnetic fields
\citep*[see, e.g.,][for a recent review]{2015SSRv..191..441H}.
In the Blandford-Payne (BP) mechanism \citep*[][]{1982MNRAS.199..883B},
a small fraction of gas in the accretion disc is accelerated by the magnetic field lines co-rotating with
the disc. {The final velocity of the driven gas is only a few times the rotational velocity of the disc \citep*[e.g.,][]
{2006MNRAS.365.1131P}, which implies that the BP mechanism is likely responsible for accelerating outflows. In the Blandford-Znajek (BZ) mechanism, the jets are powered by spinning BHs \citep*[][]{1977MNRAS.179..433B}, which is realized by the numerical simulations \citep*[e.g.,][]{2001MNRAS.326L..41K,2005MNRAS.359..801K,2004ApJ...611..977M,2005ApJ...620..878D}. It is found that magnetic forces play a key role in jet acceleration \citep*[e.g.,][]{2011MNRAS.418L..79T,2012MNRAS.423.3083M}.}
%The relative importance of the powers from the disc and BH spin remains
%under debate
%\citep*[e.g.,][]{1999ApJ...512..100L,2004ApJ...611..977M,2005ApJ...620..878D,2009ApJ...699..400G,}.
%{In the inner region of the ADAF, the magnetic pressure is
%comparable with the gas pressure. The ADAF can be magnetically
%arrested under certain circumstances \citep*[see][for the
%details]{2011ApJ...737...94C}. However, the accretion flow is
%significantly pressured in the vertical direction by the magnetic
%fields, and therefore its gas pressure can be much higher than that
%in the ADAF without magnetic fields \citep*[see the detailed results
%in][]{2011ApJ...737...94C}. This means that the magne
%c field near
%the black hole is much stronger than the estimate by assuming
%equipartition between magnetic and gas pressure with the
%conventional ADAF model.}
Other mechanisms that do not rely on
relativistic discs or BH spins may also operate
\citep*[e.g.,][]{2009MNRAS.395.2183Y}. Regardless of the details,
large-scale magnetic fields is a key ingredient of all jet
production mechanisms. Such large-scale fields are unlikely to be a
result of local MHD dynamo process, even for geometrically thick
discs \citep*[see, e.g., the simulations by
][]{2012MNRAS.423.3083M}. Thus it is natural to expect that jet
formation requires efficient advection of large-scale magnetic
fields in the disc from large distances to the inner region
\citep*[e.g.,][]{2013ApJ...764L..24S,2013ApJ...765...62S,2016ApJ...833...30C}.

The origin of large-scale magnetic fields in BH accretion discs is
directly linked to the issue of mass feeding in the outer discs. As
the external large-scale poloidal fields (e.g., the fields in the
interstellar medium in AGNs, or the fields of companion stars in
XRBs) are dragged inward by the plasma in the discs, the fields in
the inner regions of the discs can be significantly enhanced
\citep*[][]{1974Ap&SS..28...45B,1976Ap&SS..42..401B}. However, the
enhancement of magnetic fields is reduced by the magnetic diffusion
\citep*[e.g.,][]{1989ASSL..156...99V,1994MNRAS.267..235L,2001ApJ...553..158O}.
For isotropic turbulence, the effective magnetic Prandtl number
$P_{\rm m}=\nu/\eta\sim 1$ ($\nu$ is the viscosity, and $\eta$ is
magnetic diffusivity) is expected \citep*[][]{1979cmft.book.....P},
which is consistent with numerical simulations
\citep*[e.g.,][]{2003A&A...411..321Y,2009A&A...504..309L,2009A&A...507...19F,2009ApJ...697.1901G}.
It was found that field advection in a conventional
turbulence-driven thin disc is rather inefficient due to its low
radial velocity \citep*[][]{1994MNRAS.267..235L}. There are several
possible ways to resolve this difficulty of field advection in thin
discs{\citep*[e.g.,][]{2005ApJ...629..960S,2009ApJ...701..885L,2009ApJ...707..428B,2013ApJ...765..149C,2017arXiv170104627Z}}.
\citet{2005ApJ...629..960S} suggested that large-scale magnetic
fields threading the disc could be in the form of discrete,
asymmetric patches, which can efficiently remove disc angular
momentum through magnetic winds, leading to more efficient inward
drift of the field. Lovelace et al.~(2009) (see also Guilet \&
Ogilvie 2012,2013) suggested that the radial velocity of the gas in
the disc upper layer could be larger than that in the disc midplane,
which may also help field advection in thin discs.
\citet{2013ApJ...765..149C} constructed a disc-outflow model, in
which the angular momentum of the disc removed by a magnetically
accelerated outflow dominates over that caused by turbulence in the
disc. The radial velocity of the disc is therefore significantly
increased, and a weak external magnetic field can be captured to
form a strongly magnetized inner disc.

In this work, we explore the difference of jet production between
XRBs and AGNs. As noted above, the presence of large-scale magnetic fields
in the inner disc is a necessary condition for jet formation.
We compare the differences
of mass/field supplies in the outer disc boundaries and field advection through
accretion discs in these two types of
sources. We suggest that jets are switched off in the thermal state of
XRBs due to inefficient field advection in thin discs.

%%%%%%%%%%%%%%%%%%%%%%%%%%%%%%%%%%%%%%%%%%%%%%%%%%%%%%
\section{Model}\label{model}

As noted in Section 1, observations are consistent with the picture
that jet production is similar for XRBs and AGNs accreting at low
rates, while it is different at high rates -- Jets are switched off in
the thermal state of XRBs, but are observed in luminous AGNs accreting
at high rates (thermal state).  Jets can be magnetically driven by a
spinning BH (BZ mechanism) or/and from the inner region of the
accretion discs (BP mechanism).  Thus, jet formation in XRBs and AGNs
is closely associated with strong large-scale magnetic fields in the
inner regions of the discs surrounding BHs, i.e., a strong
magnetic field is a necessary condition for jet formation.

%In low/hard state of XRBs or the low-luminosity AGNs, the external
%weak magnetic fields of the gas from companion stars in XRBs or the
%gas in the interstellar medium (ISM) of AGNs can be efficiently
%dragged inward by geometrically thick ADAFs
%\citep*[][]{2011ApJ...737...94C}, and therefore strong fields are
%formed in the regions near the BHs, leading to jet
%production.
In the thermal state, a geometrically thin disc extends
to the inner-most region of the BH either in XRBs or AGNs. For a
pure viscous disc (i.e., angular momentum is transported only by
viscous stress), the radial velocity $v_r\propto (H/R)^2$, and thus
field advection is very inefficient for thin discs (small $H/R$). In
this work, we assume that most angular momentum of the thin disc is
removed via magnetic outflows, and the radial velocity of the disc
is significantly increased \citep*[see][for the
details]{2013ApJ...765..149C}. Strong magnetic field can be formed
near the BH (which then leads to jet production) only if the
condition for launching outflows from a thin disc is satisfied.

\subsection{Condition for magnetically driven
outflows}\label{condti_outflow}

We consider an accretion disc driven by magnetic outflows, i.e., the angular momentum of the disc is predominantly removed by the
outflow accelerated by the field lines threading the disc. {The field lines threading the disc are co-rotating with the disc, and a fraction of tenuous gas at the disc surface moves along the field line rotating with angular frequency $\Omega$, if the field line is inclined at an angle $\le 60^\circ$ to the disc surface \citep*[][]{1982MNRAS.199..883B}. In this case, the azimuthal velocity of the co-rotating gas $R\Omega$ increases with $R$ in the outflow till it moves beyond the Alfven radius (roughly). Therefore, the angular momentum of the gas in the disc is carried away by the outflows, which is equivalent to a magnetic torque exerting on the disc.} The magnetic torque exerted by the outflow per unit area of the disc
surface is
\begin{equation}
T_{\rm m}={\frac {B_zB_{\phi}^{\rm s}}{2\pi}}R,\label{t_m_1}
\end{equation}
where $B_{\phi}^{\rm s}$ is azimuthal component of the large scale
magnetic field at the upper disc surface, {which is caused by the mass loaded in the outflows}. The radial velocity of the disc
driven by the outflows is
\begin{equation}
v_r\simeq {\frac {2T_{\rm m}}{\Sigma R\Omega}}, \label{vr_0}
\end{equation}
where $\Omega\simeq \Omega_{\rm K}$ is assumed for weak field cases,
i.e., $\beta=8\pi p_{\rm gas}/B^2\gg 1$ \citep*[see Equation 11
in][]{2013ApJ...765..149C}. The mass accretion rate of the disc is
\begin{equation}
\dot{M}=-2\pi R\Sigma v_r\simeq-{\frac {4\pi T_{\rm m}}{\Omega_{\rm
K}}}=-{\frac {2B_z B_{\phi}^{\rm s}R}{\Omega_{\rm K}}},
\label{mdot_1}
\end{equation}
where Equations (\ref{t_m_1}) and (\ref{vr_0}) are used. We can
therefore derive a relation of the magnetic field strength with the
mass accretion rate,
\begin{displaymath}
B_z={\frac 1{\sqrt{2}}}\xi_{\phi}^{-1/2}\dot{M}^{1/2}\Omega_{\rm
K}^{1/2}R^{-1/2}~~~~~~~~~~~~~~~~~~~~~~~~~~~~
\end{displaymath}
\begin{equation}
=9.77\times 10^8\xi_{\phi}^{-1/2}\dot{m}^{1/2}m^{-1/2}r^{-5/4}~{\rm
Gauss}, \label{bz_1}
\end{equation}
where $\xi_\phi=-B_{\phi}^{\rm s}/B_z$. {The minus sign indicates the azimuthal component of the field being in the counter-direction of the disc rotation (for $B_z>0$).}
The dimensionless quantities are defined as
\begin{displaymath}
m={\frac {M}{M_\odot}},~~~\dot{m}={\frac {\dot{M}}{\dot{M}_{\rm
Edd}}},~~~\dot{M}_{\rm Edd}={\frac {L_{\rm
Edd}}{0.1c^2}},~~~r={\frac {R}{R_{\rm g}}},
\end{displaymath}
where $R_{\rm g}={{GM}/{c^2}}$.

Since the ratio $\xi_\phi\la 1$ {is required to avoid the instability (or reconnection) of
toroidal field \citep*[][]{1993noma.book.....B,1999ApJ...512..100L},}
the condition
\begin{equation}
B_z\ga B_z^{\rm min}=9.77\times
10^8\dot{m}^{1/2}m^{-1/2}r^{-5/4}~{\rm Gauss}, \label{bz_min1}
\end{equation}
must be satisfied for accretion discs driven by magnetic outflows.
This result is plotted in Figure \ref{b_crit_r} for different disc
parameters.

\subsection{Magnetic field advection of thin accretion discs driven by outflows}
\label{b_disc_outflow}

{The field dragged by the accretion disc is described by the induction equation \citep*[e.g., see Eq. 14 in][]{1994MNRAS.267..235L}.
In the steady state, the left side of the equation equals zero ($\partial/\partial t=0$). This implies that the sum of the field evolution due to field advection and diffusion vanishes. We note that the magnetic diffusion is sensitive to the field curvature in the disc, so the field inclination at the disc surface can be derived from the advection/diffusion balance {at all radii}. In this work, we consider the steady case,} i.e., $\tau_{\rm adv}=\tau_{\rm dif}$, which are estimated as
\begin{equation}
\tau_{\rm adv}\sim-{\frac R{v_r}}, \label{tau_adv_1}
\end{equation}
and
\begin{equation}
\tau_{\rm dif}\sim{\frac {RH\kappa_0}{\eta}}, \label{tau_dif_1}
\end{equation}
where $\eta$ is the magnetic diffusivity, and
$\kappa_0\equiv B_z/B_{r}^{\rm S}$ at the disc surface. The radial
velocity of the gas in the disc is then
\begin{equation}
v_r=-{\frac {\alpha c_{\rm s}}{\kappa_0 P_{\rm m}}}, \label{vr_1}
%v_r=-{\frac {\alpha P_{\rm m}c_{\rm s}}{\kappa_0}}, \label{vr_1}
\end{equation}
where $P_{\rm m}=\nu/\eta$, and $\nu=\alpha c_{\rm s}H$.

{The field inclination is crucial in accelerating outflows. The cold gas can be magnetically driven from the mid-plane of a Keplerian disc only if the field line is inclined with an angle $\le 60^\circ$ (i.e., $\kappa_0\le\sqrt{3}$) with respect to the mid-plane of the disc \citep*[][]{1982MNRAS.199..883B}. The situation is slightly different for more realistic cases (e.g., a sub-Keplerian disc)\citep*[see][]{2001ApJ...553..158O,2002A&A...385..289C}.} In this work, we adopt a necessary condition to drive magneto-centrifugal outflow as $\kappa_0 \la 1$ (approximately). {With Equation (\ref{vr_1}),} we derive the required radial advection velocity to be $|v_r|\gtrsim\alpha c_s/P_{\rm m}$.
% XC: I think it somewhat confusing, and it is difficult for the referee to understand this part, so I delete it.
%Such velocity can be
%provided by a combination of small-scale turbulent viscosity and large-scale outflow: $v_r=C_u v_r^{\rm vis}+v_r^{\rm flow}$, where
%$C_u$ is a factor (of order a few to $\sim$10) that arises from proper vertical average of the disc conductivity in the field
%advection equation \citep*[see][]{2013MNRAS.430..822G,2014ApJ...785..127O}, and $v_r^{\rm flow}$ is the radial velocity derived in Section
%\ref{condti_outflow}
%%%%%
For a conventional viscous accretion disc without magnetically driven outflows, the radial velocity is
\begin{equation}
v_{r}^{\rm vis}=-{\frac {3\nu}{2R}}=-{\frac {3\alpha c_{\rm
s}H}{2R}}. \label{vr_vis_1}
\end{equation}
Thus, we have
\begin{equation}
{\frac {v_r}{v_r^{\rm vis}}}={\frac {2}{3\kappa_0\tilde{H}P_{\rm m}}},
\label{ratio_vr_1}
\end{equation}
where $\tilde{H}=H/R$. Numerical simulations indicate that
Prandtl number $P_{\rm m}\sim 1$
\citep*[][]{2003A&A...411..321Y,2009A&A...507...19F,2009ApJ...697.1901G},
or $P_{\rm m}\sim 2-5$ \citep*[][]{2009A&A...504..309L}. In this
work, we conservatively adopt $P_{\rm m}=1$ in most of our
calculations. The ratio $v_r/v_r^{\rm vis}\sim 1/\tilde{H}$ for
$P_{\rm m}=1$ if the field line is inclined at 60$^\circ$ with respect to the disc
plane at the disc surface. This indicates that the radial velocity
of the disc driven by the outflows can be much higher than that of a
viscous thin disc. In principle, we can calculate the global
accretion disc structure with a suitable outflow solution, and then
the large-scale field advected by the disc can be obtained using the
method of \citet{1994MNRAS.267..235L}. For simplicity, we use the parameter $\xi_{v_r}\equiv v_r/v_r^{\rm vis}$ to describe
the dynamics of a magnetically driven accretion disc. {Assuming a weak vertical external magnetic field $B_{\rm ext}$ to be advected in a turbulent disc, the final steady field configuration can be calculated by solving the induction equation of the field if the radial velocity distribution $v_r(R)$ is specified \citep*[see][for the details]{1994MNRAS.267..235L}.} Approximate scalings of $B_z$ as
functions of $R$ for different values of $\xi_{v_r}$ are plotted in
Figure \ref{b_radius}. {The results show that the external weak field is substantially enhanced in the disc if its radial velocity is increased to $\sim 5-10$ times of that of a conventional viscous accretion disc, due to the presence of magnetically driven outflows. We note that the field strength at the outer radius of the disc is even lower than the external field strength $B_{\rm ext}$; this feature is caused by an artificial sharp cutoff of the disc at a certain radius $R_{\rm out}$, and was also present in the calculations of \citet{1994MNRAS.267..235L} (see Figure 1 in their paper). This feature has a very small effect on the field advection in the main body of the disc.}

%{We plot the vertical optical depth of a standard thin disc
%without outflows in Figure \ref{disc_tau}. The radial velocity of an accretion disc with outflows can
%ae $C_u$ times higher than that of a conventional turbulence driven
%accretion disc, and therefore the vertical optical depth of this
%disc will be $\sim 1/C_u$ of the standard thin disc accreting at the
%same mass rate. The vertical optical depth of a standard thin disc
%without outflows is $\ga 1000$ for typical discs surrounding a massive BH (see Figure \ref{disc_tau}).
%This means that the disc is optically thick even if $C_u$ is as high as
%10. The emission of such a disc with outflows in an AGN is still dominated by
%the thermal radiation.}

\subsection{Difference between AGN and XRB}\label{dif_agn_xrb}

The radial extent of discs around supermassive BHs is somehwat fuzzy. The thin disc may be truncated where the disc becomes
gravitationally unstable \citep*[e.g.,][]{1987Natur.329..810S,2003MNRAS.339..937G,2008A&A...477..419C}. {The zone of gravitational instability is determined by the Toomre parameter} \citep{1964ApJ...139.1217T,1965MNRAS.130...97G}. In luminous AGNs, the disc may extend to $R_{\rm out}\sim
10^{3-4}~R_{\rm g}$ \citep*[e.g.,][]{2012ApJ...761..109Y}, while the outer radii of the thin discs can be $\sim 10^{5}~R_{\rm g}$ for X-ray binaries \citep*[][]{1989MNRAS.238..897L}. For X-ray binaries, the disc must be smaller than the Roche lobe around the BH.

{In luminous AGNs, the physical processes governing the angular momentum transport of gas from galactic scales ($\sim$10~kpc) down to $\la$1~pc are quite uncertain \citep*[e.g.,][]{2005Natur.433..604D,2010MNRAS.407.1529H,2017ApJ...837..149G,2018ApJ...854..167G}. Various forms of gravitational torques are invoked to drive the inflow of gas in this region \citep*[see][for the detailed discussion, and the references therein]{2010MNRAS.407.1529H}.} In the region close to the BH, the angular momentum of the gas at the Bondi radius is roughly conserved until it approaches the circularization radius \citep*[e.g.,][]{2001ApJ...553..146M}, within which a thin disc is formed. The simulations of \citet{2014MNRAS.442..917B} are for viscously driven discs, which is not applicable for the disc driven by magnetic outflows considered in this work. {The typical magnetic field strength of galaxy cluster atmospheres is of order of $\sim\mu{\rm G}$ \citep*[see][and the references therein]{2002ARA&A..40..319C}, and the field could be stronger for the ISM in galaxies \citep*[][]{2006ApJ...645..186T,2012A&A...547A..56D}. In the central region of our Galaxy, the field strength of the gas can be as high
as $\sim{\rm mG}$ \citep*[][]{2007A&A...464..609H}. The Bondi radius of the BH is $R_{\rm B}\simeq 2GM/c_{\rm s}^2$. For $c_{\rm
s}\sim 300$~km/s (corresponding to gas temperature of $\sim 10^7$~K), $R_{\rm B}\sim 2\times 10^6R_{\rm g}$. If the angular velocity of gas at the Bondi radius is sufficient low, assuming $B\sim 10^{-6}$~G at $R_{\rm B}$, and the disc outer radius (as determined by gravitational instability or/and the angular velocity of gas at the Bondi radius) of $\sim2\times 10^3R_{\rm g}$, flux conservation implies that $B\simeq 1$~G at $R_{\rm out}$, roughly satisfies the minimum field strength requirement of Figure \ref{b_crit_r} \citep*[see][for the detailed calculations]{2016ApJ...833...30C}. Thus, the angular momentum of the thin disc may be predominantly removed by the magnetically driven outflows (see the discussion in Section \ref{condti_outflow}). The field gradually increases with decreasing radii, and therefore enable jets to be
produced in the inner regions of the discs in radio-loud AGNs. }

If the angular velocity of the gas is larger, the enhancement of magnetic fields is reduced in the flow from the Bondi radius to the outer edge of the disc due to a larger circularization radius. In this case, the condition for magnetically driven outflows may not be satisfied, and a standard thin disc is formed near the BH. It therefore appears as a radio-quiet AGN, while an AGN will appear as a radio-loud one only if the angular velocity of the circumnuclear gas is lower than a critical value at the Bondi radius \citep*[see the detailed discussion in][]{2016ApJ...833...30C}. If the external field is not ordered or if it is misaligned with the axis of the disc, the field enhancement may not be so efficient as estimated. In this case, a luminous AGN will also appear as a radio-quiet one. As the gas with a small angular velocity falls almost freely from the Bondi radius onto the outer radius of the disc, we suggest that the reconnection of the field in this region is inefficient. Numerical simulations show that shocks in a quasi-spherical slow-rotating flow may form very close to the BH (at tens of gravitational radii) \citep*[see][for the details]{2015MNRAS.447.1565S}. In our model, the thin discs are formed at $\sim 10^3-10^4 R_{\rm g}$ in bright AGNs, and we believe that the flows from the Bondi radius to the circularization radius are shock-free.

{Strong outflows driven magnetically from accretion discs are required in RL AGNs \citep*[][]{2016ApJ...833...30C}, which may alter the structure of the discs. We found that the disc remains optically thick even if its radial velocity becomes ten times that of a conventional viscous disc, which means that the local emission of the discs with outflows in RL AGNs are still nearly blackbody. Thus emergent spectra of accretion discs with outflows should have similar characteristics in UV/optical wavebands as those of standard thin discs, which is verified by the detailed calculations of the spectra of accretion discs with outflows \citep*[][]{2016ApJ...821..104Y}. This is consistent with similar UV/optical continuum spectra observed in both RL and RQ AGNs \citep*[e.g.,][]{1994ApJS...95....1E,2006ApJS..166..470R,2011ApJS..196....2S}.}

For X-ray binaries in thermal state, the external magnetic fields threading the gas from companion stars are advected inward by thin
accretion discs. If the external magnetic field strength is weaker than the minimal field strength required to drive outflows from the
outer region of the disc, the advection of the field would be very inefficient. We conjecture that this is the case for X-ray binaries,
and therefore no jets are observed in thermal state.
%No reliable measurements on the field strength of the gas from companion stars are available.
Most BH-XRBs have low-mass (solar-type) stellar companions. There is no measurement of magnetic fields in these companions as far as we are aware. However, it is safe to assume that they have solar-type magnetic fields of order 1 G. The physics of the magnetic field enhancement during the mass transfer from the companion star to the outer radius of the disc is complicated, and no detailed calculation is available till now. We suggest that  during mass transfer, the magnetic field is amplified somewhat. On the other hand, the required field at $10^5R_{\rm g}$ is $\sim 100$~G (see Figure \ref{b_crit_r}). So for most BH-LMXBs, outflows around $R_{\rm out}$ are unlikely driven from the discs. Therefore magnetic fields cannot efficiently be advected inward in most cases.

For Cyg X-1, there is a claimed measurement by \citet{2009arXiv0908.2719K}: 300 G at the companion star, 500~G at the outer disc boundary. {However, the field detected in this star is very near the reliable detection limit of the instrument \citep*[][]{2012A&A...538A.129B}, and thus further observations would be desirable. It is well known that Cyg X-1 is unique among BH-XRBs in that the companion is a massive star with $\sim 20~{M}_\odot$ \citep*[][]{2011ApJ...742...84O}. {{There is evidence that the mass is transferred through stellar wind instead of Roche lobe overflow in Cyg X-1, which indicates that the size of the disc in this source is likely smaller than that of the Roche lobe of the BH  \citep*[e.g.,][]{1976ApJ...204..555S,2009MNRAS.399.1633Z,2012MNRAS.424.1991C}. A small size of the disc implies higher magnetic field strength at the outer radius of the disc required to drive outflows (see Fig. \ref{b_crit_r}). The field in this source may still be too weak to drive outflows, even if the field of the companion star is indeed as high as claimed.}

As noted in Section 1, jets are observed in the low/hard state of X-ray
binaries, and it is believed that an ADAF is present in
such state.
%The external magnetic field can be advected inwards
%efficiently by the ADAF \citep*[][]{2011ApJ...737...94C}.
{As the luminosity increases, the ADAF is probably truncated at a smaller radius \citep*[e.g.,][]{2015A&A...573A.120P,2015ApJ...814...50D,2016MNRAS.458.2199B,2016ApJ...818L...5B,2018A&A...614A..79P,2018ApJ...867L...9T}, and then connects to a thin disc %with corona
in the region beyond the transition radius.
{Steady jets in this state indicate a strong magnetic field in the ADAF surrounding the BH, which implies sufficiently strong field strength at the outer radius of the ADAF. The detailed processes responsible for the field advection in the thin disc between the transition radius and outer disc radius are still unclear.}
It was suggested that the radial velocity of the hot gas above the disc can be higher than that at the midplane of the thin disk, which can increase the inward field advection efficiency \citep*[][]{2009ApJ...701..885L,2009ApJ...707..428B,2012MNRAS.424.2097G,2013MNRAS.430..822G}. In this picture, the weak field of the gas from the companion star is advected by the hot corona (hot gas) above the disc in the outer region, and the enhanced magnetic field at the transition radius is further dragged inwards by the inner ADAF. This may lead to a strong magnetic field accelerating jet near the BH even for an ADAF surrounded by a thin disc with corona in the outer region.}

{Although there is no clear observational evidence for a corona in the outer region of the disc \citep*[but also see][]{2005astro.ph..1215G,2010LNP...794...17G}, observations are also unable to rule out such an outer corona, because the X-ray emission from the outer region of the corona can be overwhelmed by the inner ADAF radiation (e.g., less than 10 per cent of total gravitational energy is released in the disc region with $R>100 R_{\rm g}$, and only a small fraction of it is radiated from the corona). Formation of the corona may be related to the disc instability \citep[][]{2001A&A...373..251D}, which is beyond the scope of this work. In the high/soft state, the spectrum is dominated by the thermal radiation, and a power-law spectral component may also be present but very weak \citep*[see, e.g.,][]{2012Sci...337..540F}.
%It implies that the corona is suppressed or very weak in thermal state though the detailed %physics is still unclear. So, the
External magnetic fields cannot be advected inwards efficiently by a thin disc or a very weak corona (if it is indeed present above the disc in thermal state). }

{Jets are probably accelerated by the magnetic field of the ADAF in low/hard state, and we can estimate how the jets are switched off when the XRB transits from the hard to soft state.}
{The ADAF will transit to a standard thin disc when the mass accretion rate rises above a critical value,  The duration of the transition from an ADAF to a thin disc is roughly the cooling timescale of the ADAF, which can be estimated by
\begin{equation}
t_{\rm cool}^{\rm ADAF}\sim {\frac {Hu}{F_{\rm rad}}}, \label{t_cool}
\end{equation}
where the radiation flux from the unit surface area
of the ADAF, $F_{\rm rad}=(1-f_{\rm adv})Q_+$ (with $Q_+$ the gravitational energy dissipation rate),
and $f_{\rm adv}$ is the fraction of the dissipated energy
advected in the flow. The internal energy of the gas is
\begin{equation}
u={\frac {3}{2}}n_{\rm i}kT_{\rm i}+{\frac {3}{2}}n_{\rm e}kT_{\rm
e}\simeq {\frac {3}{2}}n_{\rm i}kT_{\rm i}, \label{u}
\end{equation}
because the internal energy of the elections is negligible in ADAFs. Equation (\ref{t_cool}) can be re-written as
\begin{equation}
t_{\rm cool}^{\rm ADAF}\sim {\frac {Hu}{(1-f_{\rm adv})Q^+_{\rm ADAF}}}={\frac
{3}{2\alpha\Omega_{\rm K}f_{\Omega}^2(1-f_{\rm adv})}},
\label{t_cool2}
\end{equation}
where $c_{\rm s}=H\Omega_{\rm K}$, $f_\Omega=\Omega/\Omega_{\rm K}$, and we have used
\begin{equation}
Q^+_{\rm ADAF}={\frac {1}{2}}\nu\Sigma\left(R{\frac
{d\Omega}{dR}}\right)^2\simeq {\frac {1}{2}}\alpha c_{\rm
s}H\Sigma\Omega^2. \label{q_plus4}
\end{equation}

{If the outer thin disc is covered with a hot corona,} {the cooling timescale of the corona is
\begin{equation}
t_{\rm cool}^{\rm cor}\sim {\frac {H_{\rm cor}u}{F_{\rm rad}^{\rm cor}}}={\frac {3\tau_{\rm cor}kT_{\rm i}}{2\sigma_{\rm T}f_{\rm cor}Q^+_{\rm disc}}}, \label{t_cool_cor}
\end{equation}
where $\tau_{cor}$ is the optical depth of the corona in vertical direction, $\sigma_{\rm T}$ is electron scattering cross-section, $f_{\rm cor}$ is the fraction of the corona radiation power to the dissipation power in the underlaying disc, and the dissipation power is
\begin{equation}
Q^+_{\rm disc}\simeq {\frac {3GM\dot{M}}{8\pi R^3}}. \label{q_plus_disc}
\end{equation}
Thus Equation (\ref{t_cool_cor}) reduces to
\begin{equation}
t_{\rm cool}^{\rm cor}\sim 1.9\times 10^{-12}f_{\rm cor}^{-1}\tau_{\rm cor}m\dot{m}^{-1}r^2~~ {\rm day}, \label{t_cool_cor2}
\end{equation}
where the temperature of ions in the corona is assumed to be around virial temperature, $T_{\rm i}\sim T_{\rm vir}=GMm_{\rm p}/3kR$ \citep*[see][and the references therein]{2009MNRAS.394..207C}. The cooling timescale of the corona is around ten seconds at $r\equiv R/R_{\rm g}=100$ (or around ten minutes at $r=10^3$), while it can be $\sim100$ days at $r=10^5$, if typical values of the parameters, $f_{\rm cor}=0.1$, $\tau_{\rm cor}=0.5$, $m=10$, and $\dot{m}=0.01$ are adopted. This means that the corona in the inner region of the disc is suppressed soon after the hard to soft state transition, while only the corona near the outer edge of the disc may still survive after such a state transition. It is obvious that the external field cannot be dragged inwards efficiently, even if the field can be enhanced to some extent by the outer corona, because a barely thin disc is present in the inner region preventing further field enhancement. As discussed above, the X-ray emission from the outer corona should be very weak, which is consistent with the observational features of the thermal state.}

Steady jets are present in low/hard state of XRBs. After such an accretion mode transition, the strong magnetic field formed in the ADAF decays at the magnetic diffusion timescale of the thin disc, which is
\begin{equation}
t_{\rm dif,TD}\sim {\frac {R\kappa_0}{\eta}}={\frac
{P_{\rm m}\kappa_0R}{\alpha\Omega_{\rm K}H_{\rm TD}}}, \label{t_dif3}
\end{equation}
where $H_{\rm TD}$ is the scaleheight of the thin disc. For an ADAF accreting at the critical rate, $H/R\sim 1$,
$f_\Omega\la 1$, and $f_{\rm adv}\sim 0$. The inclination of the field $\kappa_0\sim 1$ at the surface of the ADAF.
The low/hard state transits to the thermal state at $\sim 0.1-0.5$ Eddington luminosity in most XRBs \citep*[e.g.,][]{2012Sci...337..540F}, and therefore $H_{\rm TD}/R\ga 0.1$ for the disc formed after the state transition. Comparison of Equations (\ref{t_cool2}) and (\ref{t_dif3}) shows that the diffusion timescale of the field in the thin disc is roughly comparable to the duration of accretion mode transition for a typical value of $P_{\rm m}\sim 1$. This indicates that strong jets will be switched off in the thermal state soon after the state transition \citep*[][]{2016ApJ...817...71C}. For bright AGNs, the field is dragged inwards by the gas falling onto the BH. The strong field near the BH is formed almost simultaneously with the accretion disc, which implies that jets can be formed soon after the appearance of a bright AGN at the center of a galaxy if the condition for magnetically driven outflows is satisfied. }

{An ADAF may probably exist relatively close to the BH in hard state \citep*[e.g.,][]{2015A&A...573A.120P,2015ApJ...814...50D,2016MNRAS.458.2199B,2016ApJ...818L...5B,2018A&A...614A..79P,2018ApJ...867L...9T}, and therefore sufficiently strong field strength at the outer radius of the ADAF is required for jet formation. The hot corona above the thin disc between the transition radius and the outer radius with relative thickness $H/R\sim 1$ may help field advection in this region. However, it seems to be inconsistent with the optically thin corona with $H/R\sim 0.1$ inferred from the observations of XRBs \citep*[see][for the details]{2005astro.ph..1215G,2010LNP...794...17G}. The corona scenario is also challenged by the observations of radio jets re-appearing almost immediately after transitions from the soft to hard state \citep*[][]{2018MNRAS.481.4513I}, because the soft-to-hard transitions start in the inner regions extending to the inner stable circular orbits, and the corona is therefore required to be generated instantly, which seems impossible. Thus, it seems that the corona scenario may not work in this case. The actual mechanism of the field advection in the accretion discs in hard state remains unknown, which would be an interesting issue for future investigation. }

%fig 1

\begin{figure}
  \includegraphics[width=0.48\textwidth]{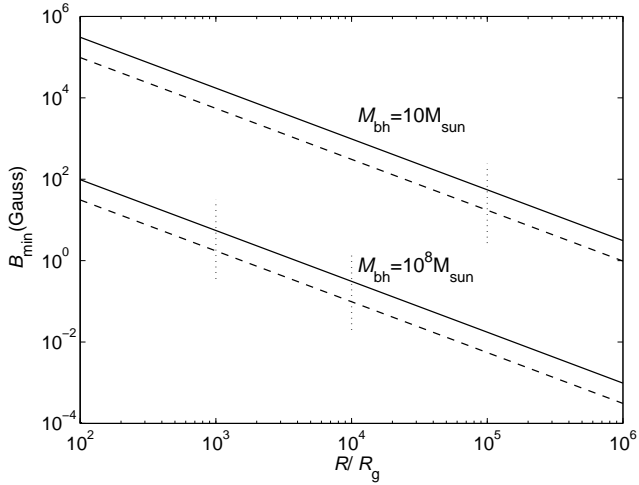}
\caption{The minimal magnetic
field strength $B_{\rm min}$ required for maintaining a magnetic
outflow driven accretion disc as functions of radius. The solid
lines represent the results calculated with $\dot{m}=0.1$, while the
dashed lines are for $\dot{m}=0.01$. The typical outer radius of the
thin discs $R_{\rm out}\sim10^3-10^4 R_{\rm g}$ for AGNs, while
$R_{\rm out}\sim 10^5 R_{\rm g}$ for XRBs (see the discussion in
Section 2.3). \label{b_crit_r} }
\end{figure}

%fig 2

\begin{figure}
  \includegraphics[width=0.48\textwidth]{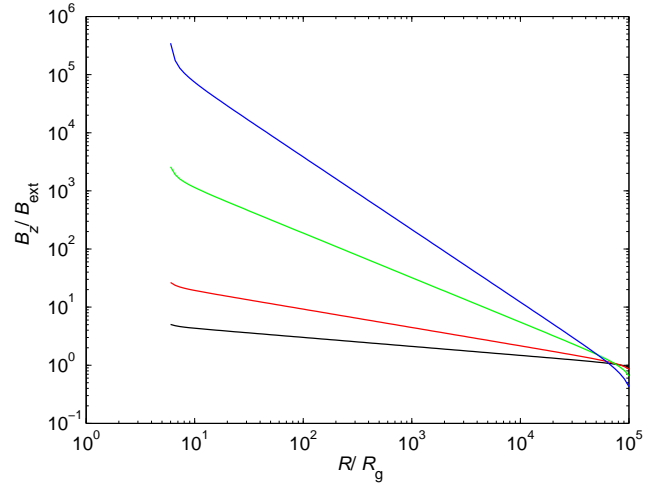}
\caption{The magnetic field
strength of the disc as functions of radius for different values of
$\xi_{v_r}\equiv v_r/v_r^{\rm vis}$, i.e., $\xi_{v_r}=1$ (black), 2
(red), 5 (green), and 10 (blue). The Prandtl number $P_{\rm m}=1$,
the outer radius of the disc $R_{\rm out}=10^5R_{\rm g}$, and the
relative disc thickness $H/R=0.1$, are adopted in all the
calculations. \label{b_radius} }
\end{figure}

%fig 3

%\begin{figure}
%  \includegraphics[width=0.48\textwidth]{disk_tau.eps}
%\caption{The true optical depth of
%the standard thin accretion discs without magnetic outflows as
%functions of radius. The effective opacity
%$\kappa_{\rm eff}=\sqrt{\kappa_{\rm R}(\kappa_{\rm es}+\kappa_{\rm R})}$ is adopted, where
%the Rosseland mean opacity $\kappa_{\rm R}=5\times 10^{24}\rho
%T^{-7/2}$~cm$^2$~g$^{-1}$. The red lines represent the
%results calculated for a BH with $M=10M_\odot$, while the
%blue lines are for a BH with $M=10^8M_\odot$. The different
%line types correspond to the discs with different parameters: solid
%($\alpha=1$ and $\dot{m}=0.1$), dashed ($\alpha=0.1$ and
%$\dot{m}=0.1$), dash-dotted ($\alpha=1$ and $\dot{m}=0.01$), and
%dotted lines ($\alpha=0.1$ and $\dot{m}=0.01$).
%\label{disc_tau} }
%\end{figure}

\section{Summary}\label{summary}

We have examined the possibility that different physical conditions
(magnetic fields in particular) in the outer discs may give rise to
qualitatively different phenomenologies of jet productions in AGNs
and XRBs. Current observations are consistent with, although do not
require, the picture that while relativistic jets are observed only
in the low/hard state and intermediate state of XRBs, they are
produced in most accretion states of AGNs, including the luminous
thermal state. We suggest that this difference arise from the
different efficiencies of magnetic field advection in AGN and XRB
discs. The production of relativistic jets requires sufficiently
strong large-scale magnetic field to build up in the inner disc, and
this in turn requires efficient field advection from the outer disc
boundary.

For geometrically thin discs that describe the thermal state of
accreting BHs, magnetic field advection is inefficient unless the
radial drift velocity in the disc is significantly larger than viscous
accretion. Magnetically driven outflow is a natual mechanism for the
enhanced radial drift, provided that the field strength is larger than
a minimum value (see Equation~\ref{bz_min1}).

In XRBs, the thin disc extends to $\sim 10^5R_{\rm g}$, and the
magnetic field strength of the gas from the companion star is
typically lower than the minimum field required to maintain an
outflow-driven accretion disc. Therefore magnetic field advection in
the thin disc of XRBs is rather inefficient, and no jets are
produced from the inner disc in the thermal state.

For radio-loud luminous AGNs, the outer radius of the thin disc is
$\sim 10^{3-4}~R_{\rm g}$. The magnetic field in the ISM is
substantially enhanced when the gas inflows from the $\sim{\rm pc}$
scale to the outer radius of the accretion disc. The magnetic field
strength at $R_{\rm out}$ is higher than the minimum strength
required to maintain an outflow-driven thin disc. Thus magnetic fields
can be efficiently advected in the thin disc, and jets can be
accelerated by the strong fields accumulated in the inner discs.

\section*{Acknowledgments}

We thank the referee for helpful comments/suggestions. Andrzej A. Zdziarski is thanked for helpful discussion. This work is supported by the NSFC (grants 11233006, 11773050 and 11833007), the CAS grant (QYZDJ-SSW-SYS023), and a visiting professorship fund from SHAO.


\begin{thebibliography}{}


\bibitem[\protect\citeauthoryear{Abramowicz et al.}{1988}]{1988ApJ...332..646A} Abramowicz M.~A., Czerny B., Lasota J.~P., Szuszkiewicz E., 1988, ApJ, 332, 646

\bibitem[\protect\citeauthoryear{Bagnulo et al.}{2012}]{2012A&A...538A.129B} Bagnulo S., Landstreet J.~D., Fossati L., Kochukhov O., 2012, A\&A, 538, A129

 \bibitem[\protect\citeauthoryear{Ballantyne \& Fabian}{2005}]{2005ApJ...622L..97B} Ballantyne D.~R., Fabian A.~C., 2005, ApJ, 622, L97

\bibitem[\protect\citeauthoryear{Basak \& Zdziarski}{2016}]{2016MNRAS.458.2199B} Basak R., Zdziarski A.~A., 2016, MNRAS, 458, 2199

 \bibitem[\protect\citeauthoryear{Beckwith, Hawley, \& Krolik}{2008}]{2008ApJ...678.1180B} Beckwith K., Hawley J.~F., Krolik J.~H., 2008, ApJ, 678, 1180-1199

\bibitem[\protect\citeauthoryear{Beckwith, Hawley, \& Krolik}{2009}]{2009ApJ...707..428B} Beckwith K., Hawley J.~F., Krolik J.~H., 2009, ApJ, 707, 428

\bibitem[\protect\citeauthoryear{Belloni}{2010}]{2010LNP...794...53B} Belloni T.~M., 2010, LNP, 794, 53

\bibitem[\protect\citeauthoryear{Belloni et al.}{2005}]{2005A&A...440..207B} Belloni T., Homan J., Casella P., van der Klis M., Nespoli E., Lewin W.~H.~G., Miller J.~M., M{\'e}ndez M., 2005, A\&A, 440, 207

\bibitem[\protect\citeauthoryear{Bernardini et al.}{2016}]{2016ApJ...818L...5B} Bernardini F., Russell D.~M., Shaw A.~W., Lewis F., Charles P.~A., Koljonen K.~I.~I., Lasota J.~P., Casares J., 2016, ApJ, 818, L5


\bibitem[\protect\citeauthoryear{Biskamp}{1993}]{1993noma.book.....B} Biskamp D., Nonlinear Magnetohydrodynamics (Cambridge: Cambridge
Univ. Press)

\bibitem[\protect\citeauthoryear{Bisnovatyi-Kogan \& Ruzmaikin}{1974}]{1974Ap&SS..28...45B} Bisnovatyi-Kogan G.~S., Ruzmaikin A.~A., 1974, Ap\&SS, 28, 45

\bibitem[\protect\citeauthoryear{Bisnovatyi-Kogan \& Ruzmaikin}{1976}]{1976Ap&SS..42..401B} Bisnovatyi-Kogan G.~S., Ruzmaikin A.~A., 1976, Ap\&SS, 42, 401

\bibitem[\protect\citeauthoryear{Blandford \& Payne}{1982}]{1982MNRAS.199..883B} Blandford R.~D., Payne D.~G., 1982, MNRAS, 199, 883

\bibitem[\protect\citeauthoryear{Blandford \& Znajek}{1977}]{1977MNRAS.179..433B} Blandford R.~D., Znajek R.~L., 1977, MNRAS, 179, 433

\bibitem[\protect\citeauthoryear{Bu \& Yuan}{2014}]{2014MNRAS.442..917B} Bu D.-F., Yuan F., 2014, MNRAS, 442, 917

\bibitem[\protect\citeauthoryear{Calderone et al.}{2013}]{2013MNRAS.431..210C} Calderone G., Ghisellini G., Colpi M., Dotti M., 2013, MNRAS, 431, 210

\bibitem[\protect\citeauthoryear{Cao}{2009}]{2009MNRAS.394..207C} Cao X., 2009, MNRAS, 394, 207

\bibitem[\protect\citeauthoryear{Cao}{2011}]{2011ApJ...737...94C} Cao X., 2011, ApJ, 737, 94

\bibitem[\protect\citeauthoryear{Cao}{2016a}]{2016ApJ...817...71C} Cao X., 2016, ApJ, 817, 71

\bibitem[\protect\citeauthoryear{Cao}{2016b}]{2016ApJ...833...30C} Cao X., 2016, ApJ, 833, 30

\bibitem[\protect\citeauthoryear{Cao \& Spruit}{2002}]{2002A&A...385..289C} Cao X., Spruit H.~C., 2002, A\&A, 385, 289

\bibitem[\protect\citeauthoryear{Cao \& Spruit}{2013}]{2013ApJ...765..149C} Cao X., Spruit H.~C., 2013, ApJ, 765, 149

\bibitem[\protect\citeauthoryear{Carilli \& Taylor}{2002}]{2002ARA&A..40..319C} Carilli C.~L., Taylor G.~B., 2002, ARA\&A, 40, 319

\bibitem[\protect\citeauthoryear{Colbert \& Mushotzky}{1999}]{1999ApJ...519...89C} Colbert E.~J.~M., Mushotzky R.~F., 1999, ApJ, 519, 89

\bibitem[\protect\citeauthoryear{Collin \& Zahn}{2008}]{2008A&A...477..419C} Collin S., Zahn J.-P., 2008, A\&A, 477, 419

\bibitem[\protect\citeauthoryear{Corbel et al.}{2001}]{2001ApJ...554...43C} Corbel S., et al., 2001, ApJ, 554, 43

\bibitem[\protect\citeauthoryear{Coriat, Fender, \& Dubus}{2012}]{2012MNRAS.424.1991C} Coriat M., Fender R.~P., Dubus G., 2012, MNRAS, 424, 1991

\bibitem[\protect\citeauthoryear{de Gasperin et al.}{2012}]{2012A&A...547A..56D} de Gasperin F., et al., 2012, A\&A, 547, A56

\bibitem[\protect\citeauthoryear{De Marco et al.}{2015}]{2015ApJ...814...50D} De Marco B., Ponti G., Mu{\~n}oz-Darias T., Nandra K., 2015, ApJ, 814, 50

\bibitem[\protect\citeauthoryear{Di Matteo, Springel, \& Hernquist}{2005}]{2005Natur.433..604D} Di Matteo T., Springel V., Hernquist L., 2005, Natur, 433, 604

\bibitem[\protect\citeauthoryear{De Villiers et al.}{2005}]{2005ApJ...620..878D} De Villiers J.-P., Hawley J.~F., Krolik J.~H., Hirose S., 2005, ApJ, 620, 878

\bibitem[\protect\citeauthoryear{Doi et al.}{2007}]{2007PASJ...59..703D} Doi A., et al., 2007, PASJ, 59, 703

\bibitem[\protect\citeauthoryear{Done \& Gierli{\'n}ski}{2005}]{2005MNRAS.364..208D} Done C., Gierli{\'n}ski M., 2005, MNRAS, 364, 208

\bibitem[\protect\citeauthoryear{Dubus, Hameury, \& Lasota}{2001}]{2001A&A...373..251D} Dubus G., Hameury J.-M., Lasota J.-P., 2001, A\&A, 373, 251

\bibitem[\protect\citeauthoryear{Duro et al.}{2016}]{2016A&A...589A..14D} Duro R., et al., 2016, A\&A, 589, A14

\bibitem[\protect\citeauthoryear{Elvis et al.}{1994}]{1994ApJS...95....1E} Elvis M., et al., 1994, ApJS, 95, 1


\bibitem[\protect\citeauthoryear{Esin, McClintock, \& Narayan}{1997}]{1997ApJ...489..865E} Esin A.~A., McClintock J.~E., Narayan R., 1997, ApJ, 489, 865

\bibitem[\protect\citeauthoryear{Fanaroff \& Riley}{1974}]{1974MNRAS.167P..31F} Fanaroff B.~L., Riley J.~M., 1974, MNRAS, 167, 31P

\bibitem[\protect\citeauthoryear{Fender}{2010}]{2010LNP...794..115F} Fender R., 2010, Lecture Notes in Physics, Berlin Springer Verlag, 794, 115

\bibitem[\protect\citeauthoryear{Fender et al.}{1999}]{1999ApJ...519L.165F} Fender R., et al., 1999, ApJ, 519, L165

\bibitem[\protect\citeauthoryear{Fender \& Belloni}{2004}]{2004ARA&A..42..317F} Fender R., Belloni T., 2004, ARA\&A, 42, 317

\bibitem[\protect\citeauthoryear{Fender \& Belloni}{2012}]{2012Sci...337..540F} Fender R., Belloni T., 2012, Sci, 337, 540

\bibitem[\protect\citeauthoryear{Fender, Belloni, \& Gallo}{2004}]{2004MNRAS.355.1105F} Fender R.~P., Belloni T.~M., Gallo E., 2004, MNRAS, 355, 1105

\bibitem[\protect\citeauthoryear{Fender, Gallo, \& Jonker}{2003}]{2003MNRAS.343L..99F} Fender R.~P., Gallo E., Jonker P.~G., 2003, MNRAS, 343, L99

\bibitem[\protect\citeauthoryear{Fromang \& Stone}{2009}]{2009A&A...507...19F} Fromang S., Stone J.~M., 2009, A\&A, 507, 19

\bibitem[\protect\citeauthoryear{Gallo, Fender, \& Pooley}{2003}]{2003MNRAS.344...60G} Gallo E., Fender R.~P., Pooley G.~G., 2003, MNRAS, 344, 60

\bibitem[\protect\citeauthoryear{Gaspari et al.}{2018}]{2018ApJ...854..167G} Gaspari M., et al., 2018, ApJ, 854, 167

\bibitem[\protect\citeauthoryear{Gaspari \& S{\c a}dowski}{2017}]{2017ApJ...837..149G} Gaspari M., S{\c a}dowski A., 2017, ApJ, 837, 149

\bibitem[\protect\citeauthoryear{Ghisellini \& Celotti}{2001}]{2001A&A...379L...1G} Ghisellini G., Celotti A., 2001, A\&A, 379, L1

\bibitem[\protect\citeauthoryear{Gilfanov}{2010}]{2010LNP...794...17G} Gilfanov M., 2010, LNP, 794, 17

\bibitem[\protect\citeauthoryear{Gilfanov \& Arefiev}{2005}]{2005astro.ph..1215G} Gilfanov M., Arefiev V., 2005, astro, arXiv:astro-ph/0501215

\bibitem[\protect\citeauthoryear{Goldreich \& Lynden-Bell}{1965}]{1965MNRAS.130...97G} Goldreich P., Lynden-Bell D., 1965, MNRAS, 130, 97

\bibitem[\protect\citeauthoryear{Goodman}{2003}]{2003MNRAS.339..937G} Goodman J., 2003, MNRAS, 339, 937

\bibitem[\protect\citeauthoryear{Gu \& Chen}{2010}]{2010AJ....139.2612G} Gu M., Chen Y., 2010, AJ, 139, 2612

\bibitem[\protect\citeauthoryear{Guan \& Gammie}{2009}]{2009ApJ...697.1901G} Guan X., Gammie C.~F., 2009, ApJ, 697, 1901

\bibitem[\protect\citeauthoryear{Guilet \& Ogilvie}{2012}]{2012MNRAS.424.2097G} Guilet J., Ogilvie G.~I., 2012, MNRAS, 424, 2097

\bibitem[\protect\citeauthoryear{Guilet \& Ogilvie}{2013}]{2013MNRAS.430..822G} Guilet J., Ogilvie G.~I., 2013, MNRAS, 430, 822

\bibitem[\protect\citeauthoryear{Han \& Zhang}{2007}]{2007A&A...464..609H} Han J.~L., Zhang J.~S., 2007, A\&A, 464, 609

\bibitem[\protect\citeauthoryear{Hardcastle, Evans, \& Croston}{2007}]{2007MNRAS.376.1849H} Hardcastle M.~J., Evans D.~A., Croston J.~H., 2007, MNRAS, 376, 1849

\bibitem[\protect\citeauthoryear{Hawley et al.}{2015}]{2015SSRv..191..441H} Hawley J.~F., Fendt C., Hardcastle M., Nokhrina E., Tchekhovskoy A., 2015, Space Sci. Rev., 191, 441

\bibitem[\protect\citeauthoryear{Hawley \& Krolik}{2006}]{2006ApJ...641..103H} Hawley J.~F., Krolik J.~H., 2006, ApJ, 641, 103

 \bibitem[\protect\citeauthoryear{Ho}{2005}]{2005Ap&SS.300..219H} Ho L.~C., 2005, Ap\&SS, 300, 219

 \bibitem[\protect\citeauthoryear{Hopkins \& Quataert}{2010}]{2010MNRAS.407.1529H} Hopkins P.~F., Quataert E., 2010, MNRAS, 407, 1529

\bibitem[\protect\citeauthoryear{Islam \& Zdziarski}{2018}]{2018MNRAS.481.4513I} Islam N., Zdziarski A.~A., 2018, MNRAS, 481, 4513

\bibitem[\protect\citeauthoryear{Jester}{2005}]{2005ApJ...625..667J} Jester S., 2005, ApJ, 625, 667

\bibitem[\protect\citeauthoryear{Jolley et al.}{2009}]{2009MNRAS.400.1521J} Jolley E.~J.~D., Kuncic Z., Bicknell G.~V., Wagner S., 2009, MNRAS, 400, 1521

\bibitem[\protect\citeauthoryear{Karitskaya et al.}{2009}]{2009arXiv0908.2719K} Karitskaya E.~A., Bochkarev N.~G., Hubrig S., Gnedin Y.~N., Pogodin M.~A., Yudin R.~V., Agafonov M.~I., Sharova O.~I., 2009, arXiv, arXiv:0908.2719

\bibitem[\protect\citeauthoryear{King et al.}{2012}]{2012ApJ...746L..20K} King A.~L., et al., 2012, ApJ, 746, L20

\bibitem[\protect\citeauthoryear{Komissarov}{2001}]{2001MNRAS.326L..41K} Komissarov S.~S., 2001, MNRAS, 326, L41

\bibitem[\protect\citeauthoryear{Komissarov}{2005}]{2005MNRAS.359..801K} Komissarov S.~S., 2005, MNRAS, 359, 801

\bibitem[\protect\citeauthoryear{Komossa et al.}{2006a}]{2006ApJ...639..710K} Komossa S., Voges W., Adorf H.-M., Xu D., Mathur S., Anderson S.~F., 2006, ApJ, 639, 710

\bibitem[\protect\citeauthoryear{Komossa et al.}{2006b}]{2006AJ....132..531K} Komossa S., Voges W., Xu D., Mathur S., Adorf H.-M., Lemson G., Duschl W.~J., Grupe D., 2006, AJ, 132, 531

\bibitem[\protect\citeauthoryear{K{\"o}rding, Jester, \& Fender}{2006}]{2006MNRAS.372.1366K} K{\"o}rding E.~G., Jester S., Fender R., 2006, MNRAS, 372, 1366

\bibitem[\protect\citeauthoryear{Kylafis \& Belloni}{2015}]{2015A&A...574A.133K} Kylafis N.~D., Belloni T.~M., 2015, A\&A, 574, A133

\bibitem[\protect\citeauthoryear{Kylafis et al.}{2012}]{2012A&A...538A...5K} Kylafis N.~D., Contopoulos I., Kazanas D., Christodoulou D.~M., 2012, A\&A, 538, A5

\bibitem[\protect\citeauthoryear{Laor \& Netzer}{1989}]{1989MNRAS.238..897L} Laor A., Netzer H., 1989, MNRAS, 238, 897

\bibitem[\protect\citeauthoryear{Lesur \& Longaretti}{2009}]{2009A&A...504..309L} Lesur G., Longaretti P.-Y., 2009, A\&A, 504, 309

\bibitem[\protect\citeauthoryear{Li \& Yan}{2016}]{2016RAA....16...48L} Li S.-L., Yan Z., 2016, RAA, 16, 48

\bibitem[\protect\citeauthoryear{Livio, Ogilvie, \& Pringle}{1999}]{1999ApJ...512..100L} Livio M., Ogilvie G.~I., Pringle J.~E., 1999, ApJ, 512, 100

\bibitem[\protect\citeauthoryear{Lovelace, Rothstein, \& Bisnovatyi-Kogan}{2009}]{2009ApJ...701..885L} Lovelace R.~V.~E., Rothstein D.~M., Bisnovatyi-Kogan G.~S., 2009, ApJ, 701, 885

\bibitem[\protect\citeauthoryear{Lubow, Papaloizou, \& Pringle}{1994}]{1994MNRAS.267..235L} Lubow S.~H., Papaloizou J.~C.~B., Pringle J.~E., 1994, MNRAS, 267, 235

\bibitem[\protect\citeauthoryear{Melia, Liu, \& Coker}{2001}]{2001ApJ...553..146M} Melia F., Liu S., Coker R., 2001, ApJ, 553, 146

\bibitem[\protect\citeauthoryear{Meyer, Liu, \& Meyer-Hofmeister}{2000}]{2000A&A...354L..67M} Meyer F., Liu B.~F., Meyer-Hofmeister E., 2000, A\&A, 354, L67
\bibitem[\protect\citeauthoryear{McKinney \& Gammie}{2004}]{2004ApJ...611..977M} McKinney J.~C., Gammie C.~F., 2004, ApJ, 611, 977

\bibitem[\protect\citeauthoryear{McKinney, Tchekhovskoy, \& Blandford}{2012}]{2012MNRAS.423.3083M} McKinney J.~C., Tchekhovskoy A., Blandford R.~D., 2012, MNRAS, 423, 3083

\bibitem[\protect\citeauthoryear{McLure \& Dunlop}{2004}]{2004MNRAS.352.1390M} McLure R.~J., Dunlop J.~S., 2004, MNRAS, 352, 1390

\bibitem[\protect\citeauthoryear{Miller et al.}{2012}]{2012ApJ...759L...6M} Miller J.~M., et al., 2012, ApJ, 759, L6

\bibitem[\protect\citeauthoryear{Neilsen \& Lee}{2009}]{2009Natur.458..481N} Neilsen J., Lee J.~C., 2009, Natur, 458, 481

\bibitem[\protect\citeauthoryear{Ogilvie \& Livio}{2001}]{2001ApJ...553..158O} Ogilvie G.~I., Livio M., 2001, ApJ, 553, 158

\bibitem[\protect\citeauthoryear{Okuzumi, Takeuchi, \& Muto}{2014}]{2014ApJ...785..127O} Okuzumi S., Takeuchi T., Muto T., 2014, ApJ, 785, 127

\bibitem[\protect\citeauthoryear{Orosz et al.}{2011}]{2011ApJ...742...84O} Orosz J.~A., McClintock J.~E., Aufdenberg J.~P., Remillard R.~A., Reid M.~J., Narayan R., Gou L., 2011, ApJ, 742, 84

\bibitem[\protect\citeauthoryear{Parker}{1979}]{1979cmft.book.....P} Parker E.~N., 1979, in Cosmical Magnetic Fields (Oxford: Clarendon), chap. 17

\bibitem[\protect\citeauthoryear{Plant et al.}{2015}]{2015A&A...573A.120P} Plant D.~S., Fender R.~P., Ponti G., Mu{\~n}oz-Darias T., Coriat M., 2015, A\&A, 573, A120

\bibitem[\protect\citeauthoryear{Poutanen, Veledina, \& Zdziarski}{2018}]{2018A&A...614A..79P} Poutanen J., Veledina A., Zdziarski A.~A., 2018, A\&A, 614, A79

\bibitem[\protect\citeauthoryear{Pudritz, Rogers, \& Ouyed}{2006}]{2006MNRAS.365.1131P} Pudritz R.~E., Rogers C.~S., Ouyed R., 2006, MNRAS, 365, 1131
\bibitem[\protect\citeauthoryear{Reis, Fabian, \& Miller}{2010}]{2010MNRAS.402..836R} Reis R.~C., Fabian A.~C., Miller J.~M., 2010, MNRAS, 402, 836

\bibitem[\protect\citeauthoryear{Richards et al.}{2006}]{2006ApJS..166..470R} Richards G.~T., et al., 2006, ApJS, 166, 470

\bibitem[\protect\citeauthoryear{Shakura \& Sunyaev}{1973}]{1973A&A....24..337S} Shakura N.~I., Sunyaev R.~A., 1973, A\&A, 24, 337

\bibitem[\protect\citeauthoryear{Shang et al.}{2011}]{2011ApJS..196....2S} Shang Z., et al., 2011, ApJS, 196, 2

\bibitem[\protect\citeauthoryear{Shapiro \& Lightman}{1976}]{1976ApJ...204..555S} Shapiro S.~L., Lightman A.~P., 1976, ApJ, 204, 555

\bibitem[\protect\citeauthoryear{Shlosman \& Begelman}{1987}]{1987Natur.329..810S} Shlosman I., Begelman M.~C., 1987, Natur, 329, 810

\bibitem[\protect\citeauthoryear{Sikora \& Begelman}{2013}]{2013ApJ...764L..24S} Sikora M., Begelman M.~C., 2013, ApJ, 764, L24

\bibitem[\protect\citeauthoryear{Sikora et al.}{2013}]{2013ApJ...765...62S} Sikora M., Stasi{\'n}ska G., Kozie{\l}-Wierzbowska D., Madejski G.~M., Asari N.~V., 2013, ApJ, 765, 62

\bibitem[\protect\citeauthoryear{Sikora, Stawarz, \& Lasota}{2007}]{2007ApJ...658..815S} Sikora M., Stawarz {\L}., Lasota J.-P., 2007, ApJ, 658, 815

\bibitem[\protect\citeauthoryear{Spruit \& Uzdensky}{2005}]{2005ApJ...629..960S} Spruit H.~C., Uzdensky D.~A., 2005, ApJ, 629, 960

\bibitem[\protect\citeauthoryear{Sukov{\'a} \& Janiuk}{2015}]{2015MNRAS.447.1565S} Sukov{\'a} P., Janiuk A., 2015, MNRAS, 447, 1565

\bibitem[\protect\citeauthoryear{Tchekhovskoy, McKinney, \& Narayan}{2008}]{2008MNRAS.388..551T} Tchekhovskoy A., McKinney J.~C., Narayan R., 2008, MNRAS, 388, 551

\bibitem[\protect\citeauthoryear{Tchekhovskoy, Narayan, \& McKinney}{2010}]{2010ApJ...711...50T} Tchekhovskoy A., Narayan R., McKinney J.~C., 2010, ApJ, 711, 50

\bibitem[\protect\citeauthoryear{Tchekhovskoy, Narayan, \& McKinney}{2011}]{2011MNRAS.418L..79T} Tchekhovskoy A., Narayan R., McKinney J.~C., 2011, MNRAS, 418, L79

\bibitem[\protect\citeauthoryear{Thompson et al.}{2006}]{2006ApJ...645..186T} Thompson T.~A., Quataert E., Waxman E., Murray N., Martin C.~L., 2006, ApJ, 645, 186

\bibitem[\protect\citeauthoryear{Toomre}{1964}]{1964ApJ...139.1217T} Toomre A., 1964, ApJ, 139, 1217

\bibitem[\protect\citeauthoryear{Tucker et al.}{2018}]{2018ApJ...867L...9T} Tucker M.~A., et al., 2018, ApJ, 867, L9

\bibitem[\protect\citeauthoryear{Ueda, Yamaoka, \& Remillard}{2009}]{2009ApJ...695..888U} Ueda Y., Yamaoka K., Remillard R., 2009, ApJ, 695, 888

\bibitem[\protect\citeauthoryear{van Ballegooijen}{1989}]{1989ASSL..156...99V} van Ballegooijen A.~A., 1989, Accretion Disks and Magnetic Fields in Astrophysics, Vol. 156, ed. G. Belvedere (Dordrecht: Kluwer), 99

\bibitem[\protect\citeauthoryear{Vasudevan \& Fabian}{2007}]{2007MNRAS.381.1235V} Vasudevan R.~V., Fabian A.~C., 2007, MNRAS, 381, 1235

\bibitem[\protect\citeauthoryear{Wu}{2009}]{2009MNRAS.398.1905W} Wu Q., 2009, MNRAS, 398, 1905

\bibitem[\protect\citeauthoryear{Wu \& Gu}{2008}]{2008ApJ...682..212W} Wu Q., Gu M., 2008, ApJ, 682, 212-217

\bibitem[\protect\citeauthoryear{Xu, Cao, \& Wu}{2009}]{2009ApJ...694L.107X} Xu Y.-D., Cao X., Wu Q., 2009, ApJ, 694, L107

\bibitem[\protect\citeauthoryear{You, Cao, \& Yuan}{2012}]{2012ApJ...761..109Y} You B., Cao X., Yuan Y.-F., 2012, ApJ, 761, 109

\bibitem[\protect\citeauthoryear{You et al.}{2016}]{2016ApJ...821..104Y} You B., Straub O., Czerny B., Sobolewska M., R{\'o}{\.z}a{\'n}ska A., Bursa M., Dov{\v c}iak M., 2016, ApJ, 821, 104

\bibitem[\protect\citeauthoryear{Yousef, Brandenburg, \& R{\"u}diger}{2003}]{2003A&A...411..321Y} Yousef T.~A., Brandenburg A., R{\"u}diger G., 2003, A\&A, 411, 321

\bibitem[\protect\citeauthoryear{Yuan et al.}{2009}]{2009MNRAS.395.2183Y} Yuan F., Lin J., Wu K., Ho L.~C., 2009, MNRAS, 395, 2183

\bibitem[\protect\citeauthoryear{Yuan et al.}{2008}]{2008ApJ...685..801Y} Yuan W., Zhou H.~Y., Komossa S., Dong X.~B., Wang T.~G., Lu H.~L., Bai J.~M., 2008, ApJ, 685, 801-827

\bibitem[\protect\citeauthoryear{Zdziarski et al.}{2004}]{2004MNRAS.351..791Z} Zdziarski A.~A., Gierli{\'n}ski M., Miko{\l}ajewska J., Wardzi{\'n}ski G., Smith D.~M., Harmon B.~A., Kitamoto S., 2004, MNRAS, 351, 791

\bibitem[\protect\citeauthoryear{Zdziarski, Kawabata, \& Mineshige}{2009}]{2009MNRAS.399.1633Z} Zdziarski A.~A., Kawabata R., Mineshige S., 2009, MNRAS, 399, 1633

\bibitem[\protect\citeauthoryear{Zdziarski, Lubi{\'n}ski, \& Smith}{1999}]{1999MNRAS.303L..11Z} Zdziarski A.~A., Lubi{\'n}ski P., Smith D.~A., 1999, MNRAS, 303, L11

\bibitem[\protect\citeauthoryear{Zhou et al.}{2003}]{2003ApJ...584..147Z} Zhou H.-Y., Wang T.-G., Dong X.-B., Zhou Y.-Y., Li C., 2003, ApJ, 584, 147

\bibitem[\protect\citeauthoryear{Zhu \& Stone}{2017}]{2017arXiv170104627Z} Zhu Z., Stone J.~M., 2017, arXiv, arXiv:1701.04627


\end{thebibliography}
\end{document}